\documentstyle[11pt]{article}

\begin{document}
%

\begin{center}
{\large \bf The Family Problem}

\vskip.5cm

W-Y. Pauchy Hwang\footnote{Correspondence Author;
 Email: wyhwang@phys.ntu.edu.tw} \\
{\em The Leung Center for Cosmology and Particle Astrophysics, \\
Institute of Astrophysics, Center for Theoretical Sciences,\\
and Department of Physics, National Taiwan University,
     Taipei 106, Taiwan}
\vskip.2cm


{\small(August 15, 2008)}
\end{center}

\begin{abstract}
We all know that in our family of particle physics we have three
generations but still don't know why - the so-called "family
problem". On other hand, in view of the masses and oscillations,
the neutrinos now present some basic difficulty in the Standard
Model. In this note, I propose that on top of the $SU_c(3)\times
SU(2) \times U(1)$ standard model there is an $SU_f(3)$ extension
- a simple $SU_c(3)\times SU(2) \times U(1) \times SU_f(3)$
extended standard model. The family gauge bosons (familons) are
massive through the so-called "colored" Higgs mechanism while the
remaining Higgs particles are also massive. The three neutrinos,
the electron-like, muon-like, and tao-like neutrinos, form the
basic family triplets. Hopefully all the couplings to the
"visible" matter are through the neutrinos, explaining why dark
matter (25 \%) is more than visible matter (5 \%) in our Universe.

\bigskip

{\parindent=0pt PACS Indices: 12.60.-i (Models beyond the standard
model); 12.10.-g (Unified field theories and models); 14.70.Pw
(Other gauge bosons).}
\end{abstract}

\section{Introduction}

More than twenty years ago I was curious by the absence of the
Higgs mechanism in the strong interactions but not it the weak
interaction sector\cite{WYPH} - a question still remains
unanswered till today. A renormalizable gauge theory that does not
have to be massless is already reputed by 't Hooft and others, for
the standard model. Maybe our question should be whether the
electromagnetism would be massless.

Another clue comes from neutrinos - they are neutral, massive and
mixing/oscillating. These particles are barely "visible" in the
Particle Table. Maybe these are avenues that connect to those
unknowns, particularly the dark matter in the Universe.

\section{A Proposal}

If we think of the role of gauge theories in quantum field theory,
we still have to recognize its unique and important role. If the
standard model is missing something, a gauge theory sector would
be one at the first guess. On the other hand, in the standard
model there are three generations of quarks and leptons. But why?
It seems to be a first loose point for the standard model. So,
let's assume that there is an $SU_f(3)$ gauge theory associated
with the story.

For this $SU_f(3)$ gauge theory, the triplet $(\nu_\tau,\,
\nu_\mu,\, \nu_e)$ serves as a basic connection. It is likely that
only the left-handed components are relevant, but we don't have to
worry in this paper. Note the up-side-down position used here
because of the way I wrote the positions of the nonzero vacuum
expectation values.

The masses of the neutrino triplet come from the coupling to some
Higgs field - a pair of complex scalar triplet, as worked out in
the previous publication\cite{WYPH}. Hereafter I ignore the
"radiative" corrections due to gauge bosons. In this case, the
eight components of the Higgs triplets are absorbed by the eight
gauge fields through the "family" Higgs mechanism via spontaneous
symmetry breaking, while the remaining four become massive Higgs
particles. (In the previous application, it was referred to
"colored Higgs Mechanism"\cite{WYPH}.)

This is the basic framework. The standard model is the gauge
theory based on the group $SU_c(3)\times SU(2)\times U(1)$. Now
the simple extension is that based on $SU_c(3)\times SU(2)\times
U(1)\times SU_f(3)$.

We may write our "new" basic elements as follows. Denote the eight
family gauge fields (familons) as $F_\mu^a(x)$. Define
$F_{\mu\nu}^a\equiv \partial_\mu F_\nu^a - \partial_\nu F_\mu^a +
\kappa f_{abc} F_\mu^b F_\nu^c$. Then we have\cite{TYWu}
\begin{equation}
{\cal L} = - {1 \over 4} F_{\mu\nu}^a F_{\mu\nu}^a.
\end{equation}
One way to describe the nonabelian nature of the gauge theory is
to add the Fadde'ev-Popov ghost fields
\begin{equation}
{\cal L}_{eff}={\cal L} - \partial_\mu {\it \phi}^a(x) D_\mu {\it
\phi}^a (x),
\end{equation}
with $D_\mu {\it \phi}^a\equiv \partial_\mu {\it \phi}^a + \kappa
f_{abc}F_\mu^b {\it \phi}^c$.

The neutrino triplet $\Psi(x)$ is
\begin{equation}
{\cal L}_f= - {\bar \Psi} \gamma_\mu D_\mu \Psi,
\end{equation}
with $D_\mu \equiv \partial_\mu - i {\kappa\over 2} \lambda^a
F_\mu^a(x)$. Just like a (triple) Dirac field.

The family Higgs mechanism is accomplished by a pair of complex
scalar triplets. Under $SU_f(3)$, they transform into the specific
forms in the U-gauge:
\begin{eqnarray}
\Phi^\prime_+=exp\{ i{\lambda_a\over 2}\xi_a^0 \} \{ u_+
+\rho_+,\, v_+
+\eta_+,\, 0 \},\nonumber\\
\Phi^\prime_-=exp\{ i{\lambda_a\over 2}\xi_a^0 \} \{ u_- +
\rho_-,\, v_- +\eta_-,\, 0 \}.
\end{eqnarray}
We could work out the kinetic terms:
\begin{equation}
{\cal L}_{scalar}=- (D_\mu \Phi_+)^\dag D_\mu \Phi_+ - (D_\mu
\Phi_-)^\dag D_\mu \Phi_- - V_\Phi,
\end{equation}
such that, by means of choosing,
\begin{equation}
u_+ =v_- =v\, cos\theta, \quad v_+ =-u_- = v\, sin\theta,
\end{equation}
we find, for the familons,
\begin{eqnarray}
M_1 = M_2 = M_3 = \kappa v, \quad M_8={\kappa v\over \sqrt 3},
\nonumber\\
M_{4,5,6,7} = {\kappa v\over \sqrt 2}.
\end{eqnarray}
That is, the eight gauge bosons all become massive. On the other
hand, by choosing
\begin{eqnarray}
&V_\Phi= {\mu^2\over 2} (\Phi_+^\dag\Phi_+ + \Phi_-^\dag\Phi_-)
\nonumber\\
&\qquad\qquad\qquad\qquad\qquad + {\lambda\over 4}\{
(\Phi_+^\dag\Phi_+)^2 + (\Phi_-^\dag\Phi_-)^2 + 2
(\Phi_+^\dag\Phi_-)(\Phi_-^\dag\Phi_+)\},
\end{eqnarray}
we find that the remaining four (Higgs) particles are massive
(with $\mu^2<0$, we have $v^2=-\mu^2/\lambda > 0$).

The other important point is the coupling between the neutrino
triplet and the family Higgs triplets:
\begin{equation}
\alpha {\bar\Psi} \times (\Phi_+ + \epsilon \Phi_-)\cdot \Psi,
\end{equation}
resulting a mass matrix which is off diagonal (but is perfectly
acceptable).

What is surprising about our model? There is no unwanted massless
particle - so, no disaster anticipated. It is the renormalizable
extension of the standard model idea. Coming back to the neutrino
sector, we now introduce the mass terms in a renormalizable way
(with the help from $SU_f(3)$ gauge theory) - previously a
headache problem in the old-day Standard model. Furthermore, there
is no major modification of the original Standard Model.

\section{Discussions}

Our life during the next stage seems to be rather difficult.
Neutrinos, albeit abundant, are very elusive. We use neutrinos as
the basic bridge to construct the $SU_f(3)$ gauge theory for the
family in the building blocks of matter. If the only coupling has
to go through neutrinos, then the detection (from the visible side
of the matter) would be extremely difficult. Of course, we should
look for the potential couplings to other sectors such as quarks
or charge leptons.

In the early universe, the temperature could be as high as that
for the familons such that the Universe could be populated with
these (interacting) particles - just like that for QCD. In other
words, our Universe would be full of these particles as the dark
matter - at this point, it is believed that our Universe has 25\%
in dark matter while only 5\% in visible matter.

\section*{Acknowledgments}
The Taiwan CosPA project is funded by the Ministry of Education
(89-N-FA01-1-0 up to 89-N-FA01-1-5) and the National Science
Council (NSC 96-2752-M-002-007-PAE). This research is also
supported in part as another National Science Council project (NSC
96-2112-M-002-023-MY3).

\end{document}